# Chopin: An Open Source R-language Tool to Support Spatial Analysis on Parallelizable Infrastructure

Insang Song[1]
Kyle P. Messier[2,3,*]

[1] Department of Geography, Seoul National University, Seoul 08826, Republic of Korea

[2] Predictive Toxicology Branch, National Institute of Environmental Health Sciences, Research Triangle Park, NC 27703, United States of America

[3] Biostatistics and Computational Biology Branch, National Institute of Environmental Health Sciences, Research Triangle Park, NC 27703, United States of America

*Corresponding author.

Abstract

This study introduces `chopin`, an R package that lowers the technical barriers to parallelizing geocomputation. Supporting popular R spatial-analysis libraries, `chopin` exploits parallel computing by partitioning data involved in each task. Partitioning can occur with regular grids, hierarchical units, or multiple file inputs, accommodating diverse input types and ensuring interoperability. This approach scales geospatial covariate calculations to match available processing power, from laptop computers to high-performance computers, reducing execution times proportional to the number of processing units. `chopin` is expected to benefit a broad range of research communities working with large-scale geospatial data, providing an efficient, flexible, and accessible tool for scaling geospatial computations.

Code metadata

Current code version: v0.9.2

Permanent link to code/repository used for this code version: https://github.com/ropensci/chopin

Legal code license: MIT License

Code versioning system used: Git

Software code languages, tools, and services used: R

Compilation requirements, operating environments and dependencies: R>=4.1, `terra`, `sf`, `exactextractr`, `mirai`, `future`, `parallelly`, `igraph`, `tidyverse`, `anticlust` (R packages)

If available, link to developer documentation/manual: https://docs.ropensci.org/chopin

Support email for questions: geoissong@snu.ac.kr

## 1. Motivation and significance

The growing body of literature across scientific disciplines such as environmental health, medical geography, remote sensing, and ecology utilize large spatial and spatio-temporal data. This trend is driven by the increasing availability of geospatial data and a growing recognition of the critical role that geographic context plays in modulating the impact of environmental factors on health outcomes. However, the computational burden of analyzing large-scale geospatial data remains a challenge, particularly in R, which lacks dedicated solutions for parallel spatial computation. This paper introduces an R-language based open-source tool to simplify computationally intensive spatial and spatio-temporal analyses by bridging spatial analysis and parallel computation packages.

The spatially and temporally varying nature of environmental data calls for the adaptation of geographic information systems (GIS) to capture them from geospatial data, making it popular in the health sciences. The analyses require expertise in GIS and proficiency in specialized software and terminology for efficient analysis of large data. Since geospatial data involve multiple dimensions, including coordinates and time, exposure assessment becomes computationally intensive. This demand has been augmented by the increasing volume of fine-resolution geospatial data and the finer spatial and temporal scale of exposure assessment. In particular, the fine-resolution exposure assessment is essential for understanding spatially varying health disparities [1-3]. Computational demand for feature extraction could be exponentially increasing depending on, to name a few, the number of features used, the spatial and temporal resolution of the raw datasets, and the definition of attributable range (i.e., buffer radius or polygon size) per dataset [4-5].

The parallel computing environment is virtually ubiquitous, which opened up a computational breakthrough for general users to “think big [6]”, allowing GIS communities to leverage parallel computing to process large volumes of geospatial datasets in a fast and efficient manner [7]. Theoretical and practical aspects of parallel computing in geospatial data processing and analysis have been discussed in the literature [8-9]. Earlier works in the 1990s laid foundations of parallel geoprocessing according to parallel computing architectures [10] and suggested data decomposition by data attributes [11-12]. With the popularization of big geospatial data, parallel solutions are developed in relation to new computing environments including cloud GIS [13-14] and parallel agent as a service [15]. Recent works developed parallel implementation of spatial models for specific applications including spatial index calculation [16], spatial simulation [17], areal interpolation [18] and geographically weighted regression [19]. Parallel geoprocessing solutions have been developed in raster and vector processing by leveraging the divisible units (e.g., raster cells) in data structure [20-22]. Existing solutions for large-scale geospatial data processing include ArcGIS [23], PostGIS [24], and tools on distributed file systems such as GeoSpark [25]. These solutions are capable of parallel processing, while posing challenges in licensing cost, database configuration, and Structured Query Language syntax, which require additional human and financial resources. In Python, `geowombat` [26] provides a compact and specialized solution for raster input and output as well as raster value extraction at point and polygon type vector data. This package is based on the `dask` extension of `dask_geopandas` [27], which supports native parallel processing functions. `dask_geopandas` extends `geopandas` [28] that is specialized in processing vector data (e.g., census boundaries and individual address points), such that common operations with vector and raster data require additional work to perform raster-vector overlay using `rasterio` [29] and `xarray` [30]. `PCML` [31] supports parallel processing of vector data, but it is at mature stage with little active maintenance. `cuSpatial` [32] harnesses graphic processing units through CUDA API for a handful of vector operations. For R language, no similar solutions to these examples exist for parallel geospatial processing. By bridging the gap between spatial and parallel computation packages, the package contributes to expanding the accessibility and efficiency of geospatial analysis, which emerges as a key quality of "next generation" GIS [33], for R language communities for statistical modeling and applied spatial research using geoprocessing results. The functions in the package implement data partitioning strategy based on geographic data parallelism, leveraging spatial proximity of data elements across data layers to distribute subsets of data to multiple processors.

## 2. Software description

**C**omputation of spatial data by **h**ierarchical and **o**bjective **p**artitioning of **in**puts for parallel processing (`chopin`) package provides a compact solution for parallelizing spatial computation in R [34]. Two major factors were considered in the development of the chopin package. First, the package should be user-friendly and interoperable with existing geospatial packages in R. Second, the package should provide succinct syntax for parallelization to flatten the learning curve for users with challenges with processing large amount of data. The R language has an array of established spatial data analysis packages such as `sf` [35], `terra` [36], both of which utilize low-level functions in Geospatial Data Abstraction Library (GDAL) [37] and Geometry Engine, Open Source (GEOS) [38].

R-level parallelization functions are adopted from `mirai` [39], `future` [40] and `future.apply` [41]. `exactextractr` provides a faster solution to calculate zonal statistics in consideration of partial overlap between vector segments and raster cells [42-43]. Functions for load balancing strategies were developed based on `igraph` [44-45] for utilizing minimum spanning tree to merge adjacent grids with a few points or polygons and `anticlust` to generate balanced clusters of points [46]. The intersection of different packages that are mostly based on low-level languages such as C and C++ contributes to enhance the processing while enabling the high-level language R to be the main interface for users with future based parallelization to leverage parallelizable computing environment. Combined, `chopin` bridges the ecosystem of spatial processing solutions in R and parallel computing environment by partitioning the data into the smaller data for parallel workers in the middle of software layers of parallel geographical applications [12].

## 2.1.Software architecture

The package includes fourteen user-facing functions, which are classified into four groups by roles in a parallel geoprocessing workflow. Six functions, of which three for two parallel backends are from`future` (`par_grid`, `par_hierarchy`, and `par_multirasters`) or `mirai` (the three functions with suffix `_mirai`), construct the main parallelization functions. These functions connect `future` or `mirai` parallel backends to common `sf` or `terra` functions as well as custom functions out of these after partitioning input data. A geoprocessing task defined as an R function with two main arguments `x` and `y` is parallelized by the spatial organization of input data in one of the main parallelization functions. Basic strategy for parallelization is to partition the large data into smaller subsets for parallel processing and calculation. Partitions are defined either by spatial extents or by file attributes. For the extent-based divisions, users can generate artificial grids or use the hierarchy existing in datasets to distribute computations per partition to parallel workers. Three common types of partitions were identified: dividing the entire region into regular grids (`par_grid`), using hierarchy explicitly or implicitly available in the data (`par_hierarchy`), and distributing operations across multiple raster files (`par_multirasters`), which are named extent, hierarchical, and raster file partitioning, respectively. All data partitioning strategies are subject to data parallelism, where the same operation is applied to different parts of the data that are independent from one another [47]. The extent partitioning approach builds upon regular decomposition of raster images [12] and resembles the processing extent functionality in proprietary software ArcGIS [48] or Python package `dask-geopandas` [17]. The other two approaches help users to distribute identical operations to parallel workers depending on users' data characteristics. Four functions are dedicated for artificial grid generation from the user's input. These functions serve for flexible grid configuration for `par_grid` to balance computational loads. Three convenience functions are designed to work seamlessly with the parallelization functions in `chopin`. `extract_at` is for raster-vector overlay with raw or buffered point, line, and polygon vector inputs. `summarize_aw` calculates summary statistics from one polygon input weighted by the other polygon data. `summarize_sedc` supports to summarize values measured at points in a certain range with distance decay applied [49].

Programmatic implementation of the package is based on metaprogramming practices including argument injection and function wrapping. Argument injection refers to the process of passing arguments into a function call programmatically [50]. In each `par_*` function, a generic `terra` and `sf` processing function, or other functions using `terra` or `sf` class data objects can be injected with its arguments. Besides the data objects, file paths can be directly entered in `par_*` functions, where terra functions are used in each parallel worker to read the file for further processing. S4 system in R language was adopted in most of the internal functions and `extract_at` to leverage multiple dispatch that allows multiple arguments in different classes to choose the correct function to apply [50]. This approach is optimal to manage multiple dependent packages to process various data classes. An argument matching function is provided to support the parallelization of generic spatial data processing functions besides ones in `terra` or `sf` packages to attain generalizability (Fig. 1). This allows users to use the same function with the same arguments as they would use in the non-parallelized version. With required and selective arguments in `par_*` functions, users can easily parallelize their existing geospatial processing code with minimal rewriting the code.

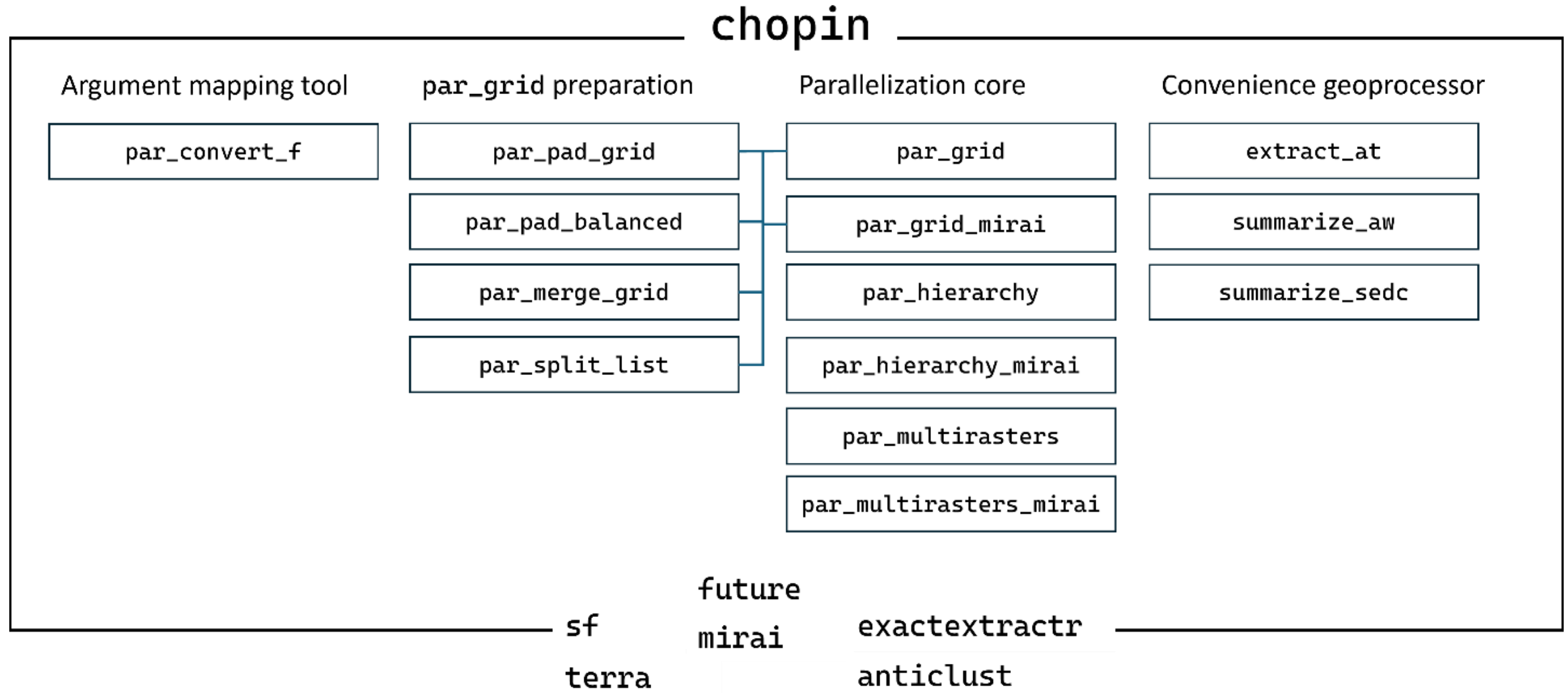

Fig. 1. Schematic representation of the `chopin` package, detailing internal functions and external dependencies.

## 2.2. Software functionalities

The standard workflow for parallel geospatial processing with `chopin` consists of running one or two functions when the user has all datasets needed to process. Users should configure parallel workers with a single line with `future` package function `plan`. When multiple threads are used, the number of threads in `workers` argument should be specified (Code Listing 1). Users may change the other parallel strategies supported in `future` package such as `cluster` and `multicore` depending on the computing environment. Other future extension solutions including `callr` plan from `future.callr` [51], `mirai_cluster` and `mirai_multisession` plans from `future.mirai` [52], and message passing interface clusters using `Rmpi` [53] could provide parallelization options including socket connections and message passing interface. Users can choose high-level configuration for parallelization by shared (`multicore` in `future`) or distributed memory (`multisession` in `future`, `mirai_multisession` in `future.mirai`, or `callr` in `future.callr`) (Code Listing 1, Table S1).

Code Listing 1. Setting up different parallel options.

```
1   # Future
2   library(future)
3   # Multisession (separate R sessions per thread)
4   plan(multisession, workers = 2)
5
6   # Multicore (forked process, only in Unix-like systems)
7   if (.Platform$OS.type != "windows") {
8     plan(multicore, workers = 2)
9   }
10
11
12  # Mirai multisession (multisession via mirai, faster than plain
    multisession)
13  library(future)
14  library(future.mirai)
15  plan(mirai_multisession, workers = 2)
16
17  # callr
```

```
18  library(future)
19  library(future.callr)
20  plan(callr, workers = 2)
21
22  # MPI cluster
23  library(future)
24  library(Rmpi)
25  library(parallelly)
26
27  cl <- makeClusterMPI(workers = 4)
28  plan(cluster, workers = cl)
```

## 2.3.Data partitioning strategies

To proceed to the description of data partitioning and load balancing, definition of terms is helpful. A *task* means a geoprocessing load assigned to the main R process (in single-threaded operation) or each parallel worker (in parallel operation). *Worker* is a cloned or spawned process of the host R process. For example, the raster value extraction using a polygon vector data is a task. Two data layers in the example are split into multiple subsets using grids, hierarchy, and raster files, and then tasks are distributed to parallel workers.

Three data partitioning strategies are implemented in `par_grid`, `par_hierarchy`, and `par_multirasters`. Extent partitioning with `par_grid` requires two steps for parallel processing, whereas the other two functions need no preparations . For extent partitioning, users first create a grid set with `par_pad_grid` or `par_pad_balanced`. `par_pad_grid` partitions the input data coverage into regular grids then conditionally performs post-processing for reducing number of grids with fewer intersecting features than a user-defined threshold. `par_pad_balanced` divides point data into equal size clusters of adjacent points then distributes each cluster to parallel worker, which results in irregular grids to subset the other input for processing. The second step is to parallelize the operation with all required inputs with `par_grid`. In other partitioning cases, `par_hierarchy` or `par_multirasters` distributes a generic operation supported in `terra` or `sf` packages or macros in `chopin`. The result object class is `data.frame` otherwise specified in the user-defined function.

For compatibility with `terra` and `sf` functions, `chopin` parallelization functions accept the fixed argument names `x` and `y` for the function that will be distributed to multiple processes. The input object class of `x` and `y` is contingent upon the processing function's specification. Users will specify which operation to parallelize and the input data according to the `sf/terra/chopin` function's argument requirements, and up to three additional arguments in each of `par_*` functions for debugging or data selection. When a function outside these packages does not support `x` and `y` arguments, users can utilize `par_convert_f` for mapping `x` and `y` to its key argument names like `par_convert_f(user_function, x = arg1, y = arg2)` when `user_function` takes two main arguments `arg1` and `arg2`. The wrapper can be used with `par_grid`, `par_hierarchy`, or `par_multirasters`.

## 2.4. Workload distribution

Table 1. Illustrations of `par_*` parallel distribution functions. (Note: red lines indicate the division of the workload.)

| Function | Strategy | Illustration | Description |
| --- | --- | --- | --- |
| `par_grid` | Extent partitioning | | Distribute workload by regular grid areas. Each parallel worker processes points in each regular grid per iteration. |
| `par_hierarchy` | Hierarchical partitioning | | Split workload hierarchically when supplementary information exists. Each parallel worker processes points in each area per iteration. |
| `par_multirasters` | Raster file partitioning | | Distribute the same task by multiple raster files. Each parallel worker processes one raster file per iteration. |

Table 2. Load-balancing strategies in extent partitioning with setting `mode` argument in `par_pad_grid` or function `par_pad_balanced`. (Note: red lines indicate the division of the workload.)

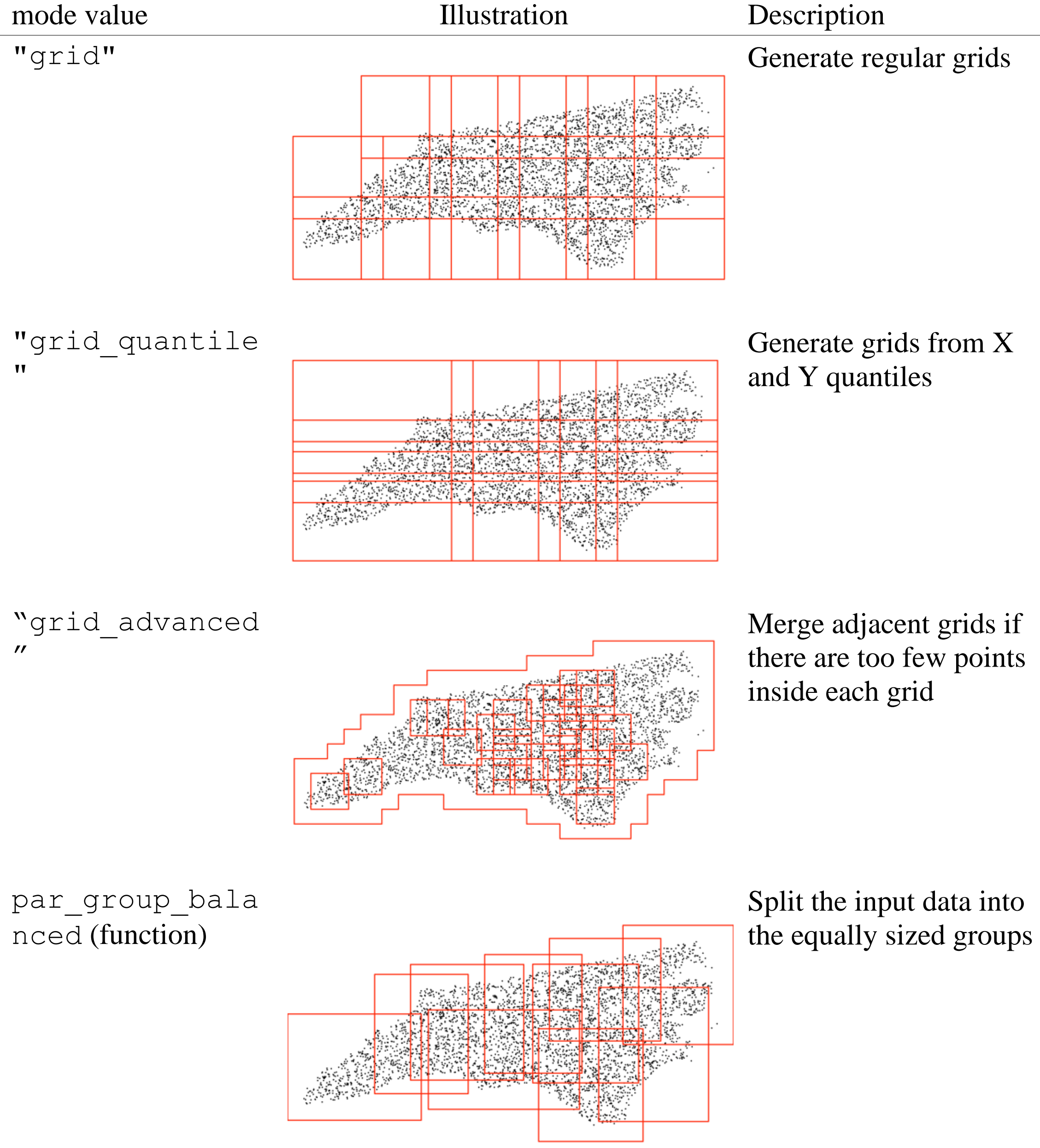

| mode value | Illustration | Description |
|---|---|---|
| `"grid"` | | Generate regular grids |
| `"grid_quantile"` | | Generate grids from X and Y quantiles |
| `“grid_advanced”` | | Merge adjacent grids if there are too few points inside each grid |
| `par_group_balanced` (function) | | Split the input data into the equally sized groups |

`par_pad_grid` is the universal helper function for two-dimensional vector data where multiple modes are supported to attain balanced loads. Two load balancing modes are supported in this function. The first mode accounts for the number of intersecting features when `mode = "grid_advanced"` is set, where in each grid to merge adjacent grids with less than the threshold feature counts. In detail, this mode uses a minimum spanning tree from adjacency matrix from the initial grids then weights the number of intersecting points in each grid. If two grids sharing a common edge both report the number of points inside them below the user-defined threshold, the grids are merged through union operation. The other mode (`mode = "grid_quantile"`) utilizes the distribution of coordinates per dimension (i.e., longitude and latitude) to obtain irregular rectangular grids from quantiles of coordinates. The created grid sets need to be passed to `par_grid` (Table 1). Grids may overlap one another depending on the value of `padding` argument. The padded grids allow to cover all raster cells or vector geometries when buffered processing extents are involved.

`par_hierarchy` splits the computation target features by the embedded hierarchy of them or an external vector object. The features in the lower hierarchy ones are grouped into the fewer and higher hierarchy. Examples include the hierarchy in Census geographies (e.g., census tracts in counties) in many countries and watersheds by stream order. Both cases will result in load balancing if computation is split along the hierarchy. To leverage `par_hierarchy`, input data are required to fulfill two conditions. First, reference and split layers include hierarchically coded values in an attribute column. Second, geometries in both features must be able to intersect one another. Intersect operation is used to obtain subsets from lower-level geometries with ones at the higher level when hierarchical relations between the two layers could not be inferred from attribute values (Table 1).

`par_multirasters` operates on multiple raster files to distribute the raster-raster or raster-vector computation to parallel workers. Each raster file is assigned to a parallel worker. This function is useful when the same operation is repeated for multiple raster files with common resolution or extent across the rasters. Calculating the summary statistics of various climate model outputs is an example. (Table 1).

Strict load balancing strategy is implemented with `par_pad_balanced`. This function divides input points into equally sized clusters based on Euclidean distance between point pairs, attaining the balanced computational loads [46]. As a result, it may equalize the computation time per point cluster in an ideal situation where the density of points is consistent over space and the target dataset has the similar resolution and density in the computation extent of each cluster.

Users can utilize coordinate quantiles per each of horizontal and vertical dimensions if the input is a point vector data (`mode = "grid_quantile"`) or merge initial regular grids with a few features inside into larger grid groups then resulting in the fewer grids with more balanced number of features per grids (`mode = "grid_advanced"`) (Table 2).

## 3. Illustrative examples

### 3.1. Average elevation extraction

An illustrative example presents `par_grid` workflow using 10,000 random points in the state of North Carolina in the United States to extract average elevation at circular point buffers. Since `par_grid` operates on a `future` parallel backend, users should define a `future` plan before running the functions. `future.mirai`'s `mirai_multisession` plan, which is supported in many platforms and generally faster than plain `future` `multisession` plan, is used below. Note that any `terra` object entered into a parallel worker is required to be converted to `sf` object. `par_grid` works with the pre-generated processing grid polygons from `par_pad_grid` function. When the input object needs to be buffered, `padding` argument value should be set larger than the buffer radius to ensure that no raster cells are omitted in the summarization. Additional arguments in `par_grid` such as `pad_y` and `.debug` are useful to designate when the parallelized operation has the required argument `x` at which `y` values are summarized. `pad_y` buffers on the processing extent of `y` is necessary to avoid omitting vector geometries or raster cells. `.debug` records error messages in the result when there is an exception while the command is run (Code Listing 2).

Code Listing 2. Extracting average elevation at counties in North Carolina with `par_grid` function in `chopin`.

```
1   library(future)
2   library(future.mirai)
3   library(sf)
4   library(terra)
5   library(chopin)
6   options(sf_use_s2 = FALSE)
```

```
7
8  # setting future.mirai parallel backend
9  future::plan(future.mirai::mirai_multisession, workers = 2L)
10
11 # read data
12 nccnty_path <- system.file(
13   "extdata", "nc_hierarchy.gpkg",
14   package = "chopin")
15 ncelev_path <-
16   system.file("extdata/nc_srtm15_otm.tif", package = "chopin")
17 nccnty <- terra::vect(nccnty_path)
18
19 ncsamp <-
20   terra::spatSample(
21     nccnty,
22     1e4L
23   )
24 ncsamp$pid <- 1:nrow(ncsamp)
25 ncsamp <- sf::st_as_sf(ncsamp)
26
27 # generate processing grids
28 ncgrid <- par_pad_grid(
29   ncsamp, mode = "grid",
30   nx = 4L, ny = 2L, padding = 10000)
31
32 # run par_grid to parallelize extract_at across grids
33 pg <-
34   par_grid(
35     grids = ncgrid,
36     fun_dist = extract_at,
37     x = ncelev_path,
38     y = ncsamp,
39     id = "pid",
40     radius = 1e4,
41     func = "mean",
42     pad_y = FALSE,
43     .debug = TRUE
44   )
```

### 3.2. Geoprocessing benchmarks

Two representative use cases were tested. The first case is set to find the distance to the nearest major road network from the United States Bureau of Transportation Statistics [54] for the Block Group population centroids in the mainland United States, obtained from the U.S. Census Bureau [55 The second case involved summarizing land use types from the National Land Use Dataset [56] at census block groups ($N = 238,193$) in the mainland United States via R package `tigris` [57]. This task requires the tabulation of the frequency of categorical raster values, which is computationally expensive. In each case, the runtime of the single-threaded computation was compared with the runtime with 32 threads. All tests were performed in a high-performance computing node at National Institute of Environmental Health Sciences (AMD 2x EPYC 9534 64-core processors and 1024GB DDR4 memory).

The results showed that parallelization reduced the computation time for both cases. The land use frequency table case took 1338.1 seconds in the single-thread mode, whereas it only took 134.1 seconds with 32 threads. Similarly, the distance to the nearest major road network case took 4427.7 seconds with a single thread, while it was completed in 84.7 seconds with 32 threads. The runtime was approximately 52.3 times faster for the distance to the nearest major road network case and 9.9 times faster for the land use frequency table case. The efficiency that was measured in the ratio of per-thread time of sequential to parallel computation was 1.634 in closest road distance calculation case, whereas 0.312 in land use frequency case (Table 3), implying that task types and underlying geoprocessing algorithm are associated with the parallel efficiency.

Table 3. Comparison of runtime and efficiency per thread in two use cases.

| Case | | | | |
|---|---|---|---|---|
| Case 1: distance to the closest road | Datasets used: primary road [54] and census block group centroids in the U.S. [55] | | | |
| | 1 thread (seconds) | 32 threads (seconds) | Parallel speedup (times) | Efficiency per thread[1] |
| | 4427.697 | 84.693 | 52.279 | 1.634 |
| Case 2: land use frequency table | Datasets used: land use classification raster (resolution: 30 meters) [56] and census block group polygon vector in the U.S. [57] | | | |
| | 1338.149 | 134.118 | 9.977 | 0.312 |

[1] (Efficiency per thread)=(Elapsed time with one thread)/((Number of threads)*(Elapsed time with multiple threads).

### 3.3. Repeated benchmarks in overlay task with multiple raster files and point vectors

We provide another set of repeated benchmarks of `par_grid` with different grid configurations and the number of threads used in averaging task against three raster layers at points ($N$=2,024,124). The same calculation was repeated for thirty times per combination of a number of grid and threads used. This case can give practical guidance in setting granularity of data parallelism and parallel efficiency of `chopin` with more CPU threads used. From the thirty repeated benchmarks, efficiency per thread decreased in log-linear trend (Fig. S1) and efficiency varied by number of cores and grid configurations (Table S2). Efficiency loss indicates that computation overhead is not trivial in parallel geospatial computation implemented in `chopin`. The reduction in parallel efficiency reiterates previous findings in parallel geoprocessing for vector queries [58] and spatial simulation [17], suggesting that the loss may be related to types of geoprocessing tasks. The results could be associated with increased clock speed under light computation loads (i.e., single thread computation) in modern CPUs and subject to Amdahl's law on the speedup limit in parallel processing [59].

### 3.4.Integrating data on cloud computers

Cloud based or distributed spatiotemporal data provision platforms such as spatiotemporal asset catalog (STAC) [60] and cloud-optimized rasters such as cloud-optimized GeoTIFF (COG) are becoming increasingly popular. The development of the alternatives may lead to efficient data processing as users are no longer required to download large data in which most parts may be unnecessary for users' focal area of analysis. The data partitioning strategy in `chopin` can take advantage of the partitioned data stream mechanism in COG to increase efficiency in reading partial data only relevant to the processing area [61]. Due to the file path centered design, `chopin` accepts the paths with proper designation of the original files as long as users have geospatial abstraction library (GDAL) supporting such formats. An example below is provided to illustrate to summarize the total population count estimates using the COG raster file uniform resource locators (URL) from OpenLandMap (https://www.openlandmap.org) STAC prefixed with the `"/vsicurl/"`. This task can also be done by `par_multirasters` leveraging multiple raster URLs (Code Listing 3).

Code Listing 3. Calculating population count from cloud-optimized GeoTIFF raster files with `chopin` directly from the internet.

```
1   library(chopin)
2   library(terra)
3   library(sf)
4   library(future.mirai)
5   library(tictoc)
6   options(sf_use_s2 = FALSE)
7
8   # URLs 2020 and 2010 population count raster
9   # with /vsicurl/ prefix
10  vsi_header <- "/vsicurl/"
11  url_arco <- "https://s3.openlandmap.org/arco/"
12  filename_p1 <- "pop.count_ghs.jrc_m_100m_s_"
13  filename_p2 <- "20%d0101_20%d1231"
14  filename_p3 <- "_go_epsg.4326_v20230620.tif"
15
16  pop10_cog <- paste0(
17    vsi_header, url_arco,
18    filename_p1, sprintf(filename_p2, 10, 10), filename_p3
19  )
20  pop20_cog <- paste0(
21    vsi_header, url_arco,
22    filename_p1, sprintf(filename_p2, 20, 20), filename_p3
23  )
24
25  urls_cog <- c(pop20_cog, pop10_cog)
26  pop_cog <- terra::rast(c(pop20_cog, pop10_cog))
27
28  # Original data from:
29  # https://data.cdc.gov/download/n44h-hy2j/application%2Fzip
30  cities500 <- "cityboundary.gpkg"
31  cities500 <- terra::vect(cities500)
32  cities500 <- terra::project(cities500, terra::crs(pop_cog))
33
34  plan(mirai_multisession, workers = 4)
35  cities500sf <- sf::st_as_sf(cities500)
36
37  cities_pop <- chopin::par_hierarchy(
38    regions = cities500sf,
39    regions_id = "ST",
40    fun_dist = extract_at,
41    x = pop20_cog,
42    y = cities500sf,
43    radius = 0,
44    id = "PLACEFIPS",
45    pad_y = FALSE,
46    .debug = TRUE
47  )
48
49  cities_pop1 <- extract_at(
```

```
50    x = pop20_cog,
51    y = cities500,
52    radius = 0,
53    id = "PLACEFIPS"
54  )
```

## 4. Discussion and Impact

`chopin` is an R-based parallel geospatial computation package for general users who want to process massive geospatial data but face challenges in adopting technical details. The package bridges the robust R package ecosystems for parallel processing and geospatial data analysis, significantly improving the usability of parallel spatial processing for a variety of use cases with spatial operations. Additionally, we provide user-friendly functions to assess spatial exposure using common operationalizations, such as extracting summary statistics at point buffers and generic polygons. This enhances the accessibility of these assessment methods to users with introductory or moderate proficiency in the R language. Our use cases demonstrate the usability and efficiency of the package in utilizing maximum computing capacity.

Two external factors may affect the efficiency of geospatial operations with `chopin`. First, the efficiency may vary depending on the operating system, as different operating systems (OS) implement different process management strategies. For example, forking for `multicore` strategy is only supported in Unix-like OS such as Linux and Mac, which support `fork()` system call [62-63], as opposed to the other `multisession` or `cluster` plan in `future` package that are also available in Windows. In these systems with `multicore` setting, the overhead of forking processes could be smaller than leveraging separated processes in `multisession` or separated physical machines supported in `cluster` plans in `future`. Second, network and storage specifications can also impact the efficiency, which usually occurs in cloud computing and cyberinfrastructures. Therefore, users are advised to consider potential bottlenecks in parallel processing when using `chopin` in distant computing resources.

As an open-source software, `chopin` is developed in a collaborative manner with assurance of reliability by adopting software development methods including continuous integration and continuous development. More than ninety one percent of the code lines were tested to ensure that the functions operate as designed. Merge rules are applied to every developer when one attempts to apply changes to the main branch, which is a term referring to a versioned copy of a code repository. This practice protects the software to be quality-controlled and robust while maintaining flexibility by accepting community contributions. External software review that is conducted via R Open Science Foundation (`https://github.com/ropensci/software-review`) provided a great opportunity to enhance the usability and align with community standards for scientific software, thereby promoting the package's adoption within the research community. The stable release of `chopin` with this paper is maintained through R-Universe repository (https://ropensci.r-universe.dev/chopin), which will be also available in the standard R package repository CRAN to make sure that the package is easily available to general users.

## 5. Conclusion

This study presented an R-based parallel geospatial computation package, `chopin`, which improves the usability and efficiency of geospatial computation with high-resolution and large-scale datasets. The package offers functions for dividing data into small processing regions to distribute large workloads over computing resources. The versatility of the package allows for optimal use of computational resources equipped with multiple CPU cores. The open-source development will promote the expansion of main functionalities to other spatial data analysis and visualization tools.

## CRediT authorship contribution statement

**Insang Song**: Conceptualization, Data curation, Formal analysis, Funding acquisition, Investigation, Methodology, Software, Writing–original draft; **Kyle P. Messier**: Funding acquisition, Supervision, Writing–review and editing.

## Declaration of competing interest

The authors declare that they have no financial interests or personal relationships that could have an influence on the work reported in this paper.

## Acknowledgements

We appreciate Dr. Eric R. Scott and Alec L. Robitaille for the thorough and comprehensive reviews on the codebase and documentation of the earlier version (0.6.3) of the package through ROpenSci software review, and Dr. Cole Brokamp, Dr. Yang Liu and Ke Xu for the constructive comments for the usability of the initial version. We appreciate Dr. Michael Fessler at the National Institute of Environmental Health Sciences for his suggestion on balancing equal loads.

## Data availability

The dataset used in the examples can be downloaded in the provided link at each code example.

## Funding

This work was supported by the Climate and Health Outcomes Research Data Systems initiative and the Intramural Research Program of the National Institutes of Health and the New Faculty Startup Fund from Seoul National University (200-20240127).

## Declaration of generative AI and AI-assisted technologies in the writing process

During the preparation of this work the authors used GitHub Copilot in order to improve the quality and legibility of `chopin` functions and examples. After using GitHub Copilot, the authors reviewed and edited the content as needed and takes responsibility for the content of the published article.

## References

[1] Parvez F, Wagstrom K. Impact of Regional Versus Local Resolution Air Quality Modeling on Particulate Matter Exposure Health Impact Assessment. Air Qual Atmos Hlth 2020;13(3):271–79. https://doi.org/10.1007/s11869-019-00786-6.

[2] Paolella DA, Tessum CW, Adams PJ, Apte JS, Chambliss S, Hill J, Muller NZ, Marshall JD. Effect of Model Spatial Resolution on Estimates of Fine Particulate Matter Exposure and Exposure Disparities in the United States. Environ Sci Technol Lett 2018;5(7):436–41. https://doi.org/10.1021/acs.estlett.8b00279.

[3] Saha PK, Presto AA, Robinson AL. Hyper-Local to Regional Exposure Contrast of Source-Resolved $PM_{2.5}$ Components Across the Contiguous United States: Implications for Health Assessment. J Expo Sci Environ Epidemiol 2023;34:836–44. https://doi.org/10.1038/s41370-023-00623-0.

[4] Ma X, Zou B, Deng J, Gao J, Longley I, Xiao S, Guo B, et al. A Comprehensive Review of the Development of Land Use Regression Approaches for Modeling Spatiotemporal Variations of Ambient Air Pollution: A Perspective from 2011 to 2023. Environ Int 2024;183:108430. https://doi.org/10.1016/j.envint.2024.108430.

[5] Clark LP, Zilber D, Schmitt C, Fargo DC, Reif DM, Motsinger-Reif AA, Messier KP. A Review of Geospatial Exposure Models and Approaches for Health Data Integration. J Expo Sci Environ Epidemiol 2024;1–18. https://doi.org/10.1038/s41370-024-00712-8.

[6] Clarke KC. Geocomputation's Future at the Extremes: High Performance Computing and Nanoclients. Parallel Comput 2003;29(10):1281–95. https://doi.org/10.1016/j.parco.2003.03.001.

[7] Yang C, Clarke K, Shekhar S, Tao CV. Big Spatiotemporal Data Analytics: A Research and Innovation Frontier. Int J Geogr Info Sci 2020;34(6):1075–88. https://doi.org/10.1080/13658816.2019.1698743.

[8] Li S, Dragicevic S, Castro FA, Sester M, Winter S, Coltekin A, Pettit C, Jiang B, Haworth J, Stein A, Cheng T. Geospatial big data handling theory and methods: A review and research challenges. ISPRS J Photogramm Remote Sens 2016;115:119–33. https://doi.org/10.1016/j.isprsjprs.2015.10.012.

[9] Werner M. Parallel Processing Strategies for Big Geospatial Data. Front Big Data 2019;2: 44. https://doi.org/10.3389/fdata.2019.00044.

[10] Ding Y, Densham PJ. Spatial strategies for parallel spatial modelling. Int J Geogr Info Sys 1996;10(6):669–98. https://doi.org/10.1080/02693799608902104.

[11] Armstrong MP, Densham PJ. Domain Decomposition for Parallel Processing of Spatial Problems. Comput Environ Urban Syst 1992;16(6):497–513. https://doi.org/10.1016/0198-9715(92)90041-O.

[12] Mineter MJ, Dowers S. Parallel Processing for Geographical Applications: A Layered Approach. J Geogr Syst 1999;1(1):61–74. https://doi.org/10.1007/s101090050005.

[13] Yang C, Goodchild M, Huang Q, Nebert D, Raskin R, Xu Y, Bambacus M, Fay D. Spatial cloud computing: how can the geospatial sciences use and help shape cloud computing? Int J Digit Earth 2011;4:305–29. https://doi.org/10.1080/17538947.2011.587547.

[14] Zhang J, Xu L, Zhang Y, Liu G, Zhao L, Wang Y. An On-Demand Scalable Model for Geographic Information System (GIS) Data Processing in a Cloud GIS. ISPRS Int J Geo-Inf 2019;8(9):392. https://doi.org/10.3390/ijgi8090392.

[15] Tan X, Guo S, Di L, Deng M, Huang F, Ye X, Sun Z, Gong W, Sha Z, Pan S. Parallel Agent-as-a-Service (P-AaaS) Based Geospatial Service in the Cloud. Remote Sens 2017;9:382. https://doi.org/10.3390/rs9040382.

[16] Bocher E, Petit G, Bernard J, Palominos S. A Geoprocessing Framework to Compute Urban Indicators: The MApUCE Tools Chain. Urban Clim 2018;24:153–74. https://doi.org/10.1016/j.uclim.2018.01.008.

[17] Kim IH, Tsou MH. Enabling Digital Earth simulation models using cloud computing or grid computing – two approaches supporting high-performance GIS simulation frameworks. Int J Digit Earth 2013;6:383–403. https://doi.org/10.1080/17538947.2013.783125.

[18] Guan Q, Kyriakidis PC, Goodchild MF. A parallel computing approach to fast geostatistical areal interpolation. Int J Geogr Info Sci 2011;25:1241–67. https://doi.org/10.1080/13658816.2011.563744.

[19] Harris R, Singleton A, Grose D, Brunsdon C, Longley P. Grid-enabling Geographically Weighted Regression: A Case Study of Participation in Higher Education in England. Trans GIS 2010;14:43–61. https://doi.org/10.1111/j.1467-9671.2009.01181.x.

[20] Lemmens R, Toxopeus B, Boerboom L, Schouwenburg M, Retsios B, Nieuwenhuis W, Mannaerts C. Implementation of a Comprehensive and Effective Geoprocessing Workflow Environment. ISPRS Ann Photogramm Remote Sens Spat Info Sci 2018;XLII-4/W8:123–27. https://doi.org/10.5194/isprs-archives-XLII-4-W8-123-2018.

[21] Shi Y, Shortridge A, Bartholic J. Grid Computing for Real Time Distributed Collaborative Geoprocessing. In: Richardson DE, van Oosterom P, editors, Advances in Spatial Data Handling. Berlin: Springer; 2002, p.197–208. https://doi.org/10.1007/978-3-642-56094-1_15.

[22] Xu C, Du X, Fan X, Giuliani G, Hu Z, Wang W, Liu J, Wang T, Yan Z, Zhu J, Jiang T, Guo H. Cloud-based storage and computing for remote sensing big data: a technical review. Int J Digit Earth 2022;15:1417–45. https://doi.org/10.1080/17538947.2022.2115567.

[23] Esri. ArcGIS Pro. Version 3.4 [software]. 2024 Nov 7.

[24] The PostGIS Development Group. 2024. PostGIS. Version 3.5.1 [software]. 2024 Dec 23. https://www.postgis.net.

[25] Gao F, Yue P. A Spark-Based Parallel Framework for Geospatial Raster Data Processing. 2018 7th International Conference on Agro-Geoinformatics (Agro-Geoinformatics) 2018;1–4. Hangzhou: IEEE. https://doi.org/10.1109/Agro-Geoinformatics.2018.8476009.

[26] GeoWombat Contributors. GeoWombat: Utilities for geospatial data. Version 2.1.22 [software]. GitHub; 2024 May 1. https://www.github.com/jgrss/geowombat.

[27] Bossche JV, Fleischmann M, Statham T, Jahn D, Augspurger T, Signell J, Gadomski, et al. Geopandas/Dask-Geopandas. Version 0.3.1 [software]. Zenodo; 2023 Apr 28. https://doi.org/10.5281/zenodo.7875807.

[28] Jordahl K, Van den Bossche J, Fleischmann M, Wasserman J, McBride J, Gerard J, Tratner J, et al. 2020. Geopandas. Version 1.0.1 [software]. Pypi; 2024 Jul 2. https://doi.org/10.5281/zenodo.3946761.

[29] Gillies S. Rasterio: Geospatial Raster I/O for Python Programmers. Version 1.4.3 [software]. Pypi; 2024 Dec 2. https://pypi.org/project/rasterio/

[30] Hoyer S, Hamman J. xarray: N-D Labeled Arrays and Datasets in Python. J Open Res Softw 2017;5(1). https://doi.org/10.5334/jors.148.

[31] Shook E, Hodgson ME, Wang S, Behzad B, Soltani K, Hiscox A, Ajayakumar J. Parallel Cartographic Modeling: A Methodology for Parallelizing Spatial Data Processing. Int J Geogr Info Sci 2016;30(12):2355–76. https://doi.org/10.1080/13658816.2016.1172714.

[32] NVIDIA Corporation. cuSpatial: GPU-Accelerated Geospatial and Spatiotemporal Algorithms. Version 25.04 [software]. NVIDIA Corporation, Santa Clara, United States; 2025 Feb 18. https://www.github.com/rapidsai/cuspatial.

[33] R Core Team. 2024. R: A Language and Environment for Statistical Computing. Version 4.4.2 [software]. R Foundation for Statistical Computing, Vienna, Austria; 2024 Oct 31. https://www.R-project.org/.

[34] Zhu AX, Zhao FH, Liang P, Qin CZ. Next generation of GIS: must be easy. Ann GIS 2021;27(1):71–86. https://doi.org/10.1080/19475683.2020.1766563.

[35] Pebesma E. Simple Features for R: Standardized Support for Spatial Vector Data. RJ 2018;10(1):439–46. https://doi.org/10.32614/RJ-2018-009.

[36] Hijmans RJ, Bivand R, Cordano E, Dyba K, Pebesma E, Sumner MD. Terra: Spatial Data Analysis. Version 1.8-5 [software]. CRAN; 2024 Dec 12. https://doi.org/10.32614/CRAN.package.terra.

[37] GDAL/OGR contributors. GDAL/OGR Geospatial Data Abstraction Software Library. Version 3.10 [software]. Zenodo; 2024 Nov 6. https://doi.org/10.5281/zenodo.5884351.

[38] GEOS contributors. GEOS Computational Geometry Library. Version 3.12.1 [software]. Zenodo; 2023 Nov 11. https://doi.org/10.5281/zenodo.11396894.

[39] Gao C, Cheng J, Hibiki AI Limited. Mirai: Minimalist Async Evaluation Framework for R. Version 1.3.1 [software]. CRAN; 2024 Nov 15. https://doi.org/10.32614/CRAN.package.mirai.

[40] Bengtsson H. A Unifying Framework for Parallel and Distributed Processing in R Using Futures. R J 2021;13(2):208–27. https://doi.org/10.32614/RJ-2021-048.

[41] Bengtsson H, R Core Team. Future.apply: Apply Function to Elements in Parallel using Futures. Version 1.11.3 [software]. CRAN; 2024 Oct 27. https://doi.org/10.32614/CRAN.package.future.apply.

[42] Baston D. exactextract. Version 0.1.0 [software]. Zenodo; 2023 Sep 12. https://doi.org/10.5281/zenodo.11396420.

[43] Baston D. exactextractr: Fast Extraction from Raster Datasets Using Polygons. Version 0.10.0 [software]. CRAN; 2023 Sep 20. https://doi.org/10.32614/CRAN.package.exactextractr.

[44] Csárdi G, Nepusz T. The igraph Software Package for Complex Network Research. InterJournal 2006;1695. https://igraph.org.

[45] Csárdi G, Nepusz T, Traag V, Horvát S, Zanini F, Noom D, Müller K. igraph: Network Analysis and Visualization in R. Version 2.1.2 [software]. CRAN; 2024 Dec 07. https://CRAN.R-project.org/package=igraph.

[46] Papenberg M, Klau GW. Using Anticlustering to Partition Data Sets into Equivalent Parts. Psychol Methods 2021;26(2):161–74. https://doi.org/10.1037/met0000301.

[47] Guan Q, Clarke KC. A General-Purpose Parallel Raster Processing Programming Library Test Application Using a Geographic Cellular Automata Model. Int J Geogr Info Sci 2010;24(5):695–722. https://doi.org/10.1080/13658810902984228.

[48] Esri. Extent (Environment Setting). https://desktop.arcgis.com/en/arcmap/latest/tools/environments/output-extent.htm; 2021 [accessed 4 January 2024].

[49] Messier KP, Akita Y, Serre ML. Integrating Address Geocoding, Land Use Regression, and Spatiotemporal Geostatistical Estimation for Groundwater Tetrachloroethylene. Environ Sci Technol 2012;46(5):2772–80. https://doi.org/10.1021/es203152a.

[50] Wickham H. Advanced R. 2nd ed. Boca Raton: CRC Press; 2019.

[51] Bengtsson H. future.callr: A Future API for Parallel Processing Using ’callr’. Version 0.8.2 [software]. CRAN; 2023 Aug 09. https://doi.org/10.32614/CRAN.package.future.callr.

[52] Bengtsson H, Gao C. future.mirai: A ’Future’ API for Parallel Processing Using ’mirai’. Version 0.2.2 [software]. CRAN; 2024 Jul 03. https://doi.org/10.32614/CRAN.package.future.mirai.

[53] Yu H. Rmpi: Interface (Wrapper) to MPI (Message-Passing Interface). Version 0.7-3.3 [software]. CRAN; 2025 Jan 13. https://doi.org/10.32614/CRAN.package.Rmpi.

[54] Bureau of Transportation Statistics. National Network Data Dictionary, https://doi.org/10.21949/1529045; 2022 [accessed 4 January 2024].

[55] U.S. Census Bureau. 2024. *Centers of Population*, https://www.census.gov/geographies/reference-files/time-series/geo/centers-population.html [accessed 4 January 2024].

[56] Dewitz J. National Land Cover Database (NLCD) 2021 Products. https://doi.org/10.5066/P9JZ7AO3; 2023 [accessed 4 January 2024].

[57] Walker K. 2015. Tigris: Load Census TIGER/Line Shapefiles. Version 2.1 [software]. CRAN; 2024 Jan 24. https://doi.org/10.32614/CRAN.package.tigris.

[58] Ilba M. Parallel algorithm for improving the performance of spatial queries in SQL: The use cases of SQLite/SpatiaLite and PostgreSQL/PostGIS databases. Comput Geosci 2021;155:104840. https://doi.org/10.1016/j.cageo.2021.104840.

[59] Amdahl GM. Validity of the single processor approach to achieving large scale computing capabilities. Proceedings of AFIPS '67 (Spring), Atlantic City, NJ. April 18-20, 1967.

[60] STAC Contributors. SpatioTemporal Asset Catalog (STAC) Specification, https://stacspec.org; 2024 [accessed 4 January 2024].

[61] Durbin C, Quinn P, Shum D. Task 51 - Cloud-Optimized Format Study. https://ntrs.nasa.gov/citations/20200001178; 2020 [accessed 4 January 2024].

[62] McCallum Q, Weston S. Parallel R: Data Analysis in the Distributed World. Sebastopol: O'Reilly Media; 2012.

[63] Baumann A, Appavoo J, Krieger O, Roscoe T. A fork() in the road. Proceedings of the Workshop on Hot Topics in Operating Systems, HotOS '19. Association for Computing Machinery, New York, United States, pp.14–22. https://doi.org/10.1145/3317550.3321435.